\documentclass[showpacs,preprintnumbers,amsmath,amssymb]{revtex4}
\usepackage{graphicx}
\usepackage{dcolumn}
\makeatletter
\input{epsf}
\begin{document}
%\preprint{APS/123-QED}
\title{Polarization of Quantum Hall States, Skyrmions and Berry Phase}
\author{B. Basu}
 \email{banasri@isical.ac.in}
\author{S. Dhar}
\email{sarmishtha_r@isical.ac.in}
\author{P. Bandyopadhyay}
 \email{pratul@isical.ac.in}
\affiliation{Physics and Applied Mathematics Unit\\
 Indian Statistical Institute\\
 KOlkata-700108 }
%\date{\today}

\begin{abstract}
We have discussed here the polarization of quantum Hall states in
the framework of the hierarchical analysis of IQHE and FQHE in
terms of Berry phase. It is observed that we have fully polarized
states for the filling factor $\nu=1$ as well as
$\nu=\frac{1}{2m+1}$, m being an integer.  However, for $\nu=p$ as
 well as $\nu =\frac{p}{q}$, with $p>1$ and odd, $q$ odd we have
partially polarized states and for $\nu=\frac{p}{q}$, $p$ even,
\end{abstract}
 \pacs{11.10.-z, 12.39.Dc, 73.43.Lp, 03.65.Vf}
\maketitle

\section{Introduction}
The quantum Hall effect (integral and fractional) appears in
two-dimensional electron systems in a strong magnetic field. In
such a high magnetic field (B $\sim 10$ T)the spin of the electron
has no dynamical role and we can assume the electrons to be
spinless in these fully polarized quantum Hall states. But when B
has relatively smaller value {\it i.e.} when the Zeeman splitting
is not so large, the corresponding system is not fully polarized.
 They are sometimes
partially polarized  and sometimes unpolarized. It is known from
experiments that the QH states at filling factors $\nu=4/3, 8/5,
....$ \cite{2,3} and at $2/3$ \cite{4,5} are unpolarized while the
states at $\nu=3/5$ \cite{5} and $7/5$ \cite{2} are partially
polarized. From numerical computations it is known \cite{6} that
the states with $\nu=\frac{2}{2n+1}$ are unpolarized and with
$\nu=\frac{1}{2n+1}$ are fully polarized in the vanishing Zeeman
splitting limit.

Wu, Dev and Jain \cite{7} have studied this problem and reported
that all even numerator QH states are unpolarized and all these
states with the numerator and denominator odd are partially
polarized or fully polarized (in the vanishing Zeeman splitting
limit). Later Mandal and Ravishankar \cite{8} proposed a global
model which accounts for all the observed quantum Hall states in
terms of an abelian doublet of Chern-Simons gauge fields, with the
strength of the Chern-Simons term given by a coupling matrix.
Hansonn, Karlhede and Leinaas \cite{9} proposed a new effective
field theory for partially polarized quantum Hall states. They
determined the density and polarization for the mean field ground
states by couplings to two Chern-Simons gauge fields. They derived
a sigma model covariantly coupled to the Chern-Simons field and
found mean field solutions which describe partially polarized
states.

It is also observed that the low energy excitations in various
polarization states are quite different. It is now confirmed that
the low energy excitation states for a fully polarized quantum
Hall state is a skyrmion whereas for unpolarized states skyrmion
excitations  are not possible. For partially polarized states
also, skyrmion excitations do not seem to occur as the skyrmionic
solitons and are found not to be of usual ones. Wu and Sondhi
\cite{10} have shown that in higher Landau levels, skyrmions are
not the low energy excitations even at small Zeeman energies. Thus
it follows that skyrmions are the quasiparticles only at $\nu=1$,
$1/3$ and $1/5$.

In some earlier papers \cite{rev11,12}, we have analyzed the
hierarchy of quantum Hall states from the view point of chiral
anomaly and Berry phase. It has been pointed out that this
approach embraces in a unified way the whole spectrum of quantum
Hall systems with their various characteristic features. Here we
propose to study the various polarization states of quantum Hall
skyrmions on the basis of this hierarchical model. Also, we shall
study the low energy excitation states for various polarization
states of quantum Hall fluid.

\section{Polarization of Integer and Fractional Quantum Hall States}
In some earlier papers \cite{rev11}, Basu and Bandyopadhyay have
studied the whole spectrum of quantum Hall states (IQH) and (FQH)
in a unified way from the chiral anomaly and Berry phase approach.
It has been shown that the hierarchy of FQH states with filling
factor $\nu=p/q$ ($p$ even or odd and $q$ odd) is interpreted in
terms of the fact that the Berry phase associated with even number
of flux quanta can be removed to the dynamical phase. For $p$ odd
and $p>1$, the corresponding state attains a higher Landau level
whereas for $p$ even, the system corresponds to particle-hole
conjugate states.

In a spherical geometry we consider quantum Hall states in the two
dimensional  surface of a 3D sphere with a magnetic monopole of
strength $\mu$ at the centre. The angular momentum relation is
given by
\begin{equation}
{\bf J} ~ = ~ {\bf r} \times {\bf p} - \mu {\bf \hat{r}} , ~ \mu
~=~ 0 , ~\pm ~1 / 2 , ~\pm ~1 , ~\pm ~ 3 / 2 \cdots.
\end{equation}
The spherical harmonics incorporating the term $\mu$ have been
extensively studied by Fierz \cite{13} and Hurst \cite{14}.
Following them we write
\begin{equation}
\begin{array}{lcl}
Y^{m, \mu}_{\ell} &=& \displaystyle{ {(1+x)}^{-(m - \mu)/2} \dot
{(1-x)}^{-(m +
\mu)/2}}\\ \\
& & \displaystyle {\frac{d^{\ell - m}}{d^{\ell - m}_x} \left[
{(1+x)}^{\ell - \mu} {(1-x)}^{\ell + \mu} \right] e^{im \phi}
e^{-i \mu \chi}}
\end{array}
\end{equation}
where $x = \cos \theta$ and the quantities $m$ and $\mu$ just
represent the eigenvalues of the operators $i
\frac{\partial}{\partial \phi}$ and $i \frac{\partial}{\partial
\chi}$ respectively.

From the description of spherical harmonics we can construct a
two-component spinor $\theta = \left( \begin{array}{c}
                          \displaystyle{u}\\
                          \displaystyle{v} \end{array} \right)$~~where
\begin{equation}
\begin{array}{lcl}
u ~=~ Y^{1/2 , 1/2}_{1/2} &=& \displaystyle{sin ~\frac{\theta}{2}
\exp
\left[ i (\phi - \chi) / 2 \right]}\\ \\
v ~=~ Y^{- 1/2 , 1/2}_{1/2} &=& \displaystyle{cos
~\frac{\theta}{2} \exp \left[- i (\phi + \chi) / 2 \right]}
\end{array}
\end{equation}
Then the $N$-particle wave function for the quantum Hall fluid can
be written as
\begin{equation}
{\psi^{(m)}}_{N} ~=~ \prod_{i < j} {(u_i v_j - u_j v_i)}^{m}
\end{equation}
where $\nu = \frac{1}{m}$, $m$ being an odd integer.

It is noted that ${\psi^{(m)}}_{N}$ is totally antisymmetric for
odd $m$ and symmetric for even $m$. Following Haldane \cite{14h},
we can identify $m$ as $m=J_i + J_j$ for an $N$-particle system
where $J_i$ is the angular momentum of the i-th particle. It is
evident from eqn. (1) that with ${\bf r} \times {\bf p}=0$ and
$\mu=\frac{1}{2}$ we have $J_i (J_j) = \frac{1}{2}$ which gives us
$m=1$, the complete filling of the lowest Landau level.
 From the Dirac quantization condition $e
\mu =\frac{1}{2}$, we note that this state corresponds to $e=1$
describing the IQH state with $\nu=1$.

The next higher angular momentum state can be achieved either by
taking ${\bf r} \times {\bf p}=1$ and $\mid\mu \mid=\frac{1}{2}$
(which implies the higher Landau level) or by taking ${\bf r}
\times {\bf p}=0$ and $\mid{\mu_{eff}} \mid=\frac{3}{2}$ implying
the ground state for the Landau level. However, with
$\mid{\mu_{eff}} \mid=\frac{3}{2}$, we find the filling fraction
$\nu=\frac{1}{3}$ which follows from the condition $e \mu
=\frac{1}{2}$ for $\mu=\frac{3}{2}$. Generalizing this we can have
$\nu=\frac{1}{5}$ with $\mid{\mu_{eff}} \mid=\frac{5}{2}$.

It is noted that when $\mu$ is an integer, we can have a relation
of the form
\begin{equation}
{\bf J} ~=~ {\bf r} \times {\bf p} - \mu {\bf \hat{r}} ~=~ {\bf
r}^{~\prime} \times {\bf p}^{~\prime}
\end{equation}
This indicates that the Berry phase which is associated with $\mu$
may be unitarily removed to the dynamical phase. Evidently, the
average magnetic field may be considered to be vanishing in these
states.
 The attachment of
$2m$ vortices to an electron effectively leads to the removal of
Berry phase to the dynamical phase. So, FQH states with  $2
\mu_{eff} = 2m + 1$ ($m$ an integer) can be viewed as if one
vortex line is attached to the {\it electron}. Now we note that
for a higher Landau level we can consider the Dirac quantization
condition $e \mu_{eff} = \frac{1}{2} n$, with $n$ being a vortex
of strength $2 \ell + 1$. This can generate FQH states having the
filling factor of the form $\frac{n}{2 \mu_{eff}}$ where both $n$
and $2 \mu_{eff}$ are odd integers. In this case, we can write $2
\mu_{eff}=2m^\prime \pm 1$ with $2m^\prime=2mn$. Indeed, we can
write the filling factor as

\begin{equation}
\nu ~=~\frac{n}{2 \mu_{eff}} ~=~ \frac{1}{\frac{2 \mu_{eff}~ \mp
1}{n} \pm \frac{1}{n}} ~=~ \frac{n}{2mn \pm 1}
\label{e1}
\end{equation}
 where $2 \mu_{eff} \mp 1$ is an even integer
given by $2m^{~\prime} = 2mn$. This is essentially the Jain
classification scheme with the constraint of $n$ being an odd
integer.

In this scheme, the FQH states with $\nu$ having the form
\begin{equation}
\nu=\frac{n^{~\prime}}{2mn^{~\prime} \pm 1}
\label{e2}
\end{equation}

with $n^\prime$ an even integer can be generated through
particle-hole conjugate states
\begin{equation}
\nu ~=~ 1 - \frac{n}{2mn \pm 1} ~=~ \frac{n (2m - 1) \pm 1}{2mn
\pm 1}=\frac{n^{~\prime}}{2mn^{~\prime} \pm 1} \label{e3}
\end{equation}
where $n(n^\prime)$ is an odd(even) integer. This leads some
observed FQH states with

\begin{equation}
\nu ~=~ \frac{p}{q}(p~{\rm even}, q~{\rm odd})~~:~~\frac{2}{3} ,
~\frac{2}{5} , ~\frac{4}{5} , ~\frac{4}{7} , ~\frac{4}{9} ,
~\frac{6}{11} , ~\frac{6}{13} , \cdots \label{e4}
\end{equation}

Besides. for $\nu = \frac{p}{q}$ with $p > q$ we can think of
these states as condensates of one IQH state with $\nu$ an integer
and one FQH state with $\nu ~=~ \frac{p}{q}$ ($p < q$). In this
category, we have even numerator states such as $\nu=~ \frac{4}{3}
, ~\frac{8}{5}$ and $~\frac{10}{7}$ \cite{2,3}.

Now to study the polarization states of various FQH systems, we
observe that in the lowest Landau level, we have the filling
factor $\nu$ given by $\nu= \frac{1}{2 \mu_{eff}} = \frac{1}{2m +
1}$. As we have pointed out that  the attachment of $2m$ vortices
to an electron leads to the removal of Berry phase to the
dynamical phase and this effectively corresponds to the attachment
of one vortex(magnetic flux) to an electron, this electron will be
a polarized one.

This is also true for $\nu=1$ state, as in this case we have
$\mu=\frac{1}{2}$ implying one vortex line (magnetic flux)is
attached to an electron.

However, in the higher Landau level, this scenario will change.
Indeed, in this case we have $\nu=\frac{n}{2m n + 1}$ with $n>1$
and an odd integer.

Now from the relation (\ref{e1}), for $n>1$ and an odd integer
\begin{equation}
\nu ~=~\frac{n}{2 \mu_{eff}} ~=~ \frac{1}{\frac{2 \mu_{eff}~ \mp
1}{n} \pm \frac{1}{n}}
\end{equation}

where $2 \mu_{eff} \mp 1$ is an even integer. We note that as even
number of flux units can be accommodated  in the dynamical phase
we may consider this as $\frac{1}{n}$ flux unit is attached to an
electron. This suggests that electrons will not be fully polarized
as one full flux unit is not available to it. This will correspond
to partially polarized states with $\nu=p/q$ ($p>1$ and odd, $q$
odd) which represents the states like
$\frac{3}{5},~\frac{3}{7},~\frac{5}{9}$ and so on. It is observed
here that IQH states like $\nu=3,5$ will also exhibit partially
polarized states as from the Dirac quantization condition $e \mu
=\frac{n}{2}$, for $n>1$ and odd, the filling factor $\nu=n$ is
achieved with $\mu=1/2$. This suggest that one flux unit is shared
by an electron in $n$ number of Landau levels so that each
electron is attached with $\frac{1}{n}$ flux unit implying a
partially polarized state.

Now we consider the states with the filling factor $\nu=p/q$ with
$p$ even and $q$ odd. As we have pointed out earlier, this is
achieved when we have particle-hole conjugate states given by
$\nu=1-\frac{n}{2m n \pm 1}$ with $n$ an odd integer. A hole
configuration is described by the complex conjugate of the
particle state, the spin orientation of the particle and hole
state will be opposite to each other. Thus this will represent an
unpolarized state. This gives us the general relation of
unpolarized FQH states with filling factor $\nu=p/q$, $p$ even and
$q$ odd.

\section{Skyrmion Excitations in Arbitrary Polarized Quantum Hall
States}

It has been found that in quantum Hall systems deviations from the
incompressible filling factor $\nu$ is accomplished by the
degradation of the system's  spin polarization. This effect has
been observed near $\nu = 1$ in several experiments that directly
probe the spin density of the electron gas \cite{15e,16e,17e,18e}
]. By noticing that the dynamics of quantum Hall system with a
spin polarized ground state will follow that of a quantum
ferromagnet and that the skyrmion is a charged object of the
system, Sondhi et.al.\cite{19,20} proposed a phenomenological
action which is valid for the long wave length and small frequency
limit. In this scheme, the competition between the Zeeman and
Coulomb terms sets the size and energy of the skyrmions. Recently,
Basu, Dhar and Bandyopadhyay \cite{21,22} have proposed a pure
sigma model formalism for skyrmions taking resort to spherical
geometry where $2D$ electron gas resides on the surface of a $3D$
sphere with a magnetic monopole placed at the centre. To study the
system in {\it pure sigma model} formalism an $O(4)$ nonlinear
sigma model in $3+1$ dimensional manifold was introduced. In this
framework, the quartic stability term introduced by Skyrme, known
as the Skyrme term, determines the size of the quantum Hall
skyrmions and the effect of the Zeeman energy and Coulomb energy
is encoded in a certain parameter such that the size and energy
are determined from the sigma model Lagrangian with the Skyrme
term. Besides, the introduction of the topological $\theta$-term
in the Lagrangian helps us to determine the spin and statistics of
the skyrmion. The four order parameter fields help us to construct
two independent $SU(2)$ algebras which are associated with the two
mutually opposite orientations of the magnetization vector which
resides on the $2D$ surface of the sphere.

The generalized Lagrangian is taken to be of the form \cite{21}

\begin{equation}
\begin{array}{lcl}
L~
&=&\displaystyle{~2\pi{{\cal{J}}}^S_\mu~{\cal{A}}_\mu-~\frac{M^2}{16}~Tr(\partial_{\mu}
U^{\dag}\partial_\nu U)-\frac{1}{32 \eta^2}~Tr[\partial_\mu
UU^{\dag},\partial_\nu UU^{\dag}]^2}\\
&&\displaystyle{-\frac{\theta}{16\pi^2}~
{^{*}{\mathcal{F}}_{\mu\nu}}{\mathcal{F}}_{\mu\nu}-\frac{1}{4}{\mathcal{F}}_{\mu\nu}{\mathcal{F}}^{\mu\nu}}\\
\end{array}
\end{equation}
Here the topological current is defined as

\begin{equation}
{{\mathcal{J}}}^S_{\mu}~=~\frac{1}{24 \pi^2}~
\epsilon_{\mu\nu\alpha\beta}
Tr(U^{-1}\partial_{\nu}U)(U^{-1}\partial_{\alpha}U)
(U^{-1}\partial_{\beta}U)
\end{equation}
where ${\cal{A}}_{\mu}$ is a four-vector gauge field,
$\theta=g/c^2$ with $g=\nu e^2/h$, the Hall conductivity and
$~^{*}{\mathcal{F}}_{\mu \nu}$   is a Hodge dual given by
\begin{equation}
^{*}{\mathcal{F}}_{\mu\nu}
=\frac{1}{2}~\epsilon_{\mu\nu\lambda\sigma}{\mathcal{F}}_{\lambda\sigma}
\end{equation}
$M$ is a constant of dimension of mass and $\eta$ is a
dimensionless coupling parameter.

 The $SU(2)$ matrix $U$ is here defined as
\begin{equation}
U=n_0 I + \bf{n}.\overrightarrow{\tau}
\end{equation}
where the chiral fields $n_0$,$n_1$,$n_2$ and $n_3$ satisfy the
relation $\sum n_i^2=1$. It may be noted that in $2+1$ dimension,
we have the normalized $3$-vector field ${\bf n}$ with $\sum
n^2_i=1$ where $n_i$ corresponds to the local spin direction. In
the $O(4)$ model $n_i (i=1,2,3)$ corresponds to this spin
direction which live on the $2$-dimensional surface of the sphere
where the extra field $n_0$ helps us to consider three {\it boost}
 generators in $(n_0, n_i)$ planes. In view of this, we can
consider two types of generators such that the generator $M_k$
rotates the $3$-vector $\bf{n}(x)$ to any chosen axis and the
boost generators $N_k$ would mix $n_0$ with the components of
$\bf{n}$. We can now construct the following algebra

$$[M_i,M_j]= {i \epsilon_{ijk} M_k}$$
$$[M_i,N_j]= {i \epsilon_{ijk} N_k}$$
\begin{equation}
[N_i,N_j]= {i \epsilon_{ijk} M_k}
\end{equation}
which is locally isomorphic  to the Lie algebra of the $O(4)$
group. This helps us to introduce the left and right generators
$$L_i=\frac{1}{2}(M_i-N_i)$$
\begin{equation}
R_i=\frac{1}{2}(M_i+N_i)
\end{equation}
which satisfy
$$[L_i,L_j]= {i \epsilon_{ijk} L_k}$$
$$[R_i,R_j]= {i \epsilon_{ijk} R_k}$$
\begin{equation}
[L_i,R_j]=0
\end{equation}

Thus the algebra has split into two independent subalgebras each
isomorphic to a $SU(2)$ algebra and corresponds to the chiral
group $SU(2)_L \otimes SU(2)_R$. The left and right chiral group
can now be taken to be associated with two mutually opposite
orientations of the magnetization vector which resides on the $2$D
surface of the sphere.

The topological charge of quantum Hall skyrmions is given by the
winding number
 \begin{equation}
  Z~=~ {\frac{1 }{24
\pi^2}}~\int_{S^3} dS_{\mu}~ \epsilon^{\mu\nu\lambda\sigma}
[(U^{-1}
\partial_{\nu} U)(U^{-1} \partial_{\lambda} U)(U^{-1} \partial_{\sigma} U)]
\end{equation}
which is related to the homotopy $\pi_3(S^3)=Z$ and the electric
charge is given by $\nu e Z$. We observe here that the Pontryagin
index given by
\begin{equation}
q=2\mu=~-{\frac{1}{16 \pi^2}}~\int Tr ^{*} {\mathcal{F}}_{\mu\nu}
{\mathcal{F}}_{\mu\nu} d^4 x
\end{equation}

introduces the Berry phase for quantum Hall states as $\mu$
represents magnetic monopole strength. When there is a monopole of
strength $\mu$ at the centre, the flux through the sphere is $2
\mu$ and the phase is given by $e^{i\phi_B}$ where $\phi_B=2 \pi
\nu$ (number of flux quanta enclosed by the loop).
 For $Z=1$, the skyrmion when moving around a closed loop
acquires the Berry phase $2\pi\nu N$ where $N$ is the number of
skyrmions enclosed by the loop. To find the spin and statistics,
we consider a process which exchanges two skyrmions in the rest
frame of the other skyrmion. This exchange effectively corresponds
to the other skyrmion moving around the first in a half circle and
hence it picks up a phase $\pi~\nu$ which is the statistical phase
of a skyrmion. For $\nu=1$ it is a fermion and $\nu=1/m$, $m$
being an odd integer it corresponds to an anyon in planar
geometry. In general, the spin of the skyrmion having charge $\nu
e Z$ is given by $\nu Z/2$.

Now to study the relevant quasiparticles for arbitrarily polarized
states we observe the following possibilities.

1)Fully Polarized States :

In this case the spin orientation is fixed : up($\uparrow$) or
down($\downarrow$). So for the chiral group $SU(2)_L \times
SU(2)_R$ only one $SU(2)_L (SU(2)_R)$ group will be operative.
From the homotopic relation $\pi_3(SU(2))=Z$, we will have
solitons (skyrmions) with charge $\nu e Z$.

2)Partially Polarized States :

In this case it will not be possible to split the $O(4)$ group
into two independent left and right chiral groups $SU(2)_L$ or
$SU(2)_R$ as each of these represent sharp polarization states.
Thus we cannot represent the relevant quasiparticles as skyrmions.

3)Unpolarized States :

In this case, we will have equal number of up and down spin
electrons. Now from the homotopic relation $\pi_3(SU(2) \times
SU(2))=\pi_3(SO(4))=0$, we note that there will be no skyrmions.

Thus we find that skyrmions are the relevant quasiparticles only
at the filling factor given by $\nu=1$ and $\nu=\frac{1}{2m+1}$
with $m=$ an integer.

\section{Discussion}
We have studied here the polarization states of quantum hall fluid
at different filling factors from the hierarchical scheme in the
framework of Berry phase. It is found that for IQH and FQH states
with $\nu=1$ and $\nu=\frac{1}{2m+1}$, $m$ being an integer
represent fully polarized states. The states with $\nu=p$ as well
as $\nu=p/q,~ p$ odd, $q$ odd with $p>1$ represent partially
polarized states and with $\nu=p/q, ~p$ even, $q$ odd with $p>1$
represent unpolarized states. It may be noted that the same result
for unpolarized states has also been found from numerical
computations \cite{6} and an exact diagonalization study has
revealed that at $\nu=3/5$, the system is partially polarized.
These are also in agreement with experiments. Mandal and
Ravishankar \cite{8} have studied the polarization states in terms
of an Abelian doublet of Chern-Simons gauge fields where the
strength of the Chern-Simons term is given by a coupling matrix in
the framework of composite fermion model. Their findings are also
consistent with these results.

Wu and Sondhi \cite{10} have calculated the energies of
quasiparticles for odd integer filling factors $\nu=2k+1,~k\geq 1
$ and have observed that skyrmions are not the low energy charged
excitations even at small Zeeman energies. Mandal and Ravishankar
\cite{23} have studied the quasiparticles in partially polarized
states and have observed that skyrmions in this case are not the
usual one whereas for unpolarized states skyrmions do not exist at
all. Here from the homotopic analysis we have found that skyrmions
are the relevant quasiparticles only for fully polarized states
whereas for partially polarized and unpolarized states skyrmionic
excitations do not exist at all.

\end{document}